\documentstyle[14pt]{article}
\begin{document}
\def\t{\tau}
\def\a{\alpha}
\def\b{\beta}
\def\I{{\rm I}}
\def\({\left(}
\def\){\right)}
\def\[{\left[}
\def\]{\right]}
\def\p{\prime}
\def\d{\dagger}
\def\bar#1{\overline #1}
\def\up#1#2{\stackrel{{}_{(#1)}}{#2}}
\def\vecr#1{\mid{#1}\rangle}
\def\vecl#1{\langle{#1}\mid}
\def\th{\theta}
\def\pfi{\varphi}
\def\s{\sigma_3}
\def\L{\Lambda}
\def\O{\Omega}
\def\G{\Gamma}
\def\R{{\rm R}}
\def\sgn{{\rm sgn}}
\def\g{\gamma}
\def\eps{\varepsilon}
\begin{center}
{\Large Perturbation theory for
the modified nonlinear Schr{\"o}dinger solitons~}
\end{center}

\begin{center}
V.S. Shchesnovich${}^a$ and E.V. Doktorov${}^b$\\
\medskip
${}^a${\small\it Division for Optical Problems in Information Technologies,
National
Academy of Sciences of Belarus, Zhodinskaya St. 1/2, 220141 Minsk,
Republic of Belarus}\\
${}^b${\small\it B.I.Stepanov Institute of Physics, 68 F. Skaryna Ave.,
220072 Minsk, Republic of Belarus}
\end{center}
\bigskip

\hrule
\medskip
\noindent{\small{\bf Abstract}}
\medskip

{\small The perturbation theory based on the Riemann-Hilbert problem is developed
for the modified nonlinear Schr{\"o}dinger equation which describes the
propagation of femtosecond optical pulses in nonlinear single-mode optical
fibers. A detailed analysis of the adiabatic approximation to
perturbation-induced evolution of the soliton parameters is given. The
linear perturbation and the Raman gain are considered as examples.}
\medskip
\hrule
\bigskip

\medskip

\bigskip

\noindent
{\bf 1. Introduction}
\medskip

The study of the dynamical processes associated with the propagation of
high-power optical pulses in single-mode nonlinear fibers is based as a
rule on the integrable nonlinear Schr{\"o}dinger equation (NLSE)~[1].
Various realistic effects accompanying the soliton propagation and
destroying the integrability of the NLSE are usually treated as
perturbations. There are different approaches to describe analytically
perturbation-induced dynamics of the NLS solitons~[2-8], for review see
Ref.~9. It is evident that the "quality" of taking into account for the
above effects depends crucially on smallness of a parameter responsible
for a definite perturbation. Just this situation takes place with the Kerr
nonlinearity dispersion effect. Being sufficiently small in the picosecond
pulse duration region, it becomes essential for femtosecond solitons,
having a parameter of the order $10^{-2}-10^{-1}$. Hence, streactly
speaking, this effect cannot be treated as a perturbation for the
femtosecond region of soliton pulse duration.

The natural approach to treat analytically the dynamics of femtosecond
solitons is to consider the so-called perturbed modified nonlinear Schr{\"o}dinger
equation (MNLSE)~[10]
\begin{equation}
iq_z+\frac{1}{2}q_{\t\t}+i\a\(|q|^2q\)_\t+|q|^2q=r,
\end{equation}
where the term with the real parameter $\a$ governs the effect of the Kerr
nonlinearity dispersion (self-steepening) and $r$ accounts for small
effects which we will consider as perturbation. Here $q(\t,z)$ is the
normalized slowly varying amplitude of the complex field envelope, $z$ is
the normalized propagation distance along the fiber, $\t$ is the normalized
time measured in a frame of reference moving with the pulse at the group
velocity (the retarded time). It is remarkable that MNLSE~(1) with zero
r.h.s. is still integrable by the inverse scattering transform (IST)
method~[11], though the linear spectral problem associated with the MNLSE is
different from that for the NLSE.

Our primary goal is to develop a
simple formalism to treat analytically the femtosecond soliton dynamics
governed by Eq.~(1). Three points should be stressed which differ our
approach from the previously known ones. First, we account for the Kerr
nonlinearity dispersion effect {\it exactly}. In other words, we do
not make any hypothesis about smallness of~$\a$ in Eq.~(1). Moreover,
we consider as a background solution not the sech-like
pulse of the NLS type but precisely the MNLS soliton. Finally, the third
point is relevant to the formalism, namely, we employ the Riemann-Hilbert
(RH) problem, which was proved to be effective for treating perturbations
to nonlinear evolution equations integrable by means of the Zakharov-Shabat
spectral problem~[12-15]. Recently, we developed the RH problem-based
approach [16-18] for solving nonlinear equations integrable by the
Wadati-Konno-Ichikawa spectral problem~[11]. This approach includes the
MNLSE and some its generalizations. This development serves as a base for
taking into account small perturbations.

The paper is arranged as follows. In Sec.~2 the basic results on the RH
problem-based approach to the MNLSE is summarized. In Sec.~3 the general
one-soliton solution of the MNLSE is derived in a form which, we believe,
is as simple as possible. Here we also discuss the limiting transition to
the NLS soliton. In Sec.~4 we obtain the perturbation-induced evolution
equations for the RH problem data related to the soliton parameters and
discuss peculiarities of the perturbation theory for gauge equivalent
equations. Sec.~5 is devoted to the adiabatic approximation. Here we
consider as an example the linear perturbation (excess gain or fiber loss)
and the Raman self-frequency shift~[19-21]. Concluding remarks are contained
in the last section.
\bigskip

\noindent
{\bf 2. Riemann-Hilbert problem for MNLSE}
\medskip

In this section we summarize the basic results concerning the approach to
the MNLSE based on the RH probem. Let us write the MNLSE
in the general form
$$
iq_z+\frac{1}{2}q_{\t\t}+i\a\(|q|^2q\)_\t+\b|q|^2q=0,
\eqno(2)$$
where $\a$ and~$\b$ are real parameters.
Eq.~(2) is integrable via the IST method and can be represented as the
compatibility condition $U_z-V_\t+[U,V]=0$ for the following system of two
linear matrix equations
$$
\Phi_\t=\L(k)[\s,\Phi]+2ikQ\Phi\equiv U\Phi-\L(k)\Phi\s,
\qquad\qquad\qquad\qquad\qquad\eqno(3a)$$
$$
\Phi_z=\O(k)[\s,\Phi]+\(\frac{4i}{\a}k^3Q+2ik^2Q^2\s-\frac{i\b}{\a}kQ
+kQ_\t\s-2i\a kQ^3\)\Phi
$$
$$
\equiv V\Phi-\O(k)\Phi\s.
\eqno(3b)$$
Here the Hermitian matrix $Q=\(\begin{array}{cc}0&q\\\bar q&0\end{array}\)$
represents the potential of the linear spectral problem~(3$a$),
$\L(k)=-({2i}/{\a})\(k^2-{\b}/{4}\)$,
$\O(k)=-({4i}/{\a^2})(k^2$ $-{\b}/{4})^2$, the bar stands for complex
conjugation and $k$ is a spectral parameter.

As seen from Eq.~(3$a$), the MNLSE~(2) belongs to the class of equations
integrable by means of the Wadati-Konno-Ichikawa spectral problem~[11].
However, Eq.~(2) is not canonical among the equations of this class~[16]. The
canonical equation
$$
iq^\p_z+\frac{1}{2}q^\p_{\t\t}-i\a {q^\p}^2\bar q^\p_\t+\b|q^\p|^2q^\p
+\a^2|q^\p|^4q^\p=0
\eqno(4)$$
does not admit as obvious a physical interpretation as the MNLSE, but
possesses the Lax representation too,
$$
\Phi^\p_\t=\L(k)[\s,\Phi^\p]+\(2ikQ^\p+i\a {Q^\p}^2\s\)\Phi^\p
\equiv U^\p\Phi^\p-\L(k)\Phi^\p\s,
\quad\quad\eqno(5a)$$
$$
\Phi^\p_z=\O(k)[\s,\Phi^\p]+\Biggl(\frac{4i}{\a}k^3Q^\p+2ik^2{Q^\p}^2\s
-\frac{i\b}{\a}kQ^\p+kQ^\p_\t\s +\frac{\a}{2}[Q^\p,Q^\p_\t]
$$
$$
\qquad\quad+\frac{i\a^2}{2}{Q^\p}^4\s\Biggr)\Phi^\p
\equiv V^\p\Phi^\p-\O(k)\Phi^\p\s,
\eqno(5b)$$
where
$Q^\p=\(\begin{array}{cc}0&q^\p\\  \bar q^\p&0\end{array}\)$. The
spectral problem~(5$a$) associated with Eq.~(4), as distinct from the
spectral problem~(3$a$), is compatible with the canonical normalization
condition $\Phi^\p(k=\infty)=\I$, where I is the $2\times2$ identity
matrix. Eqs.~(2) and~(4) are gauge equivalent equations interrelated by the
following  gauge transformation:
$$
Q=g^{-1}Q^\p g,
\eqno(6)$$
where $g(\t,z)=\Phi^\p(k=0)$. Thereby, solutions of the MNLSE can be
obtained from those of Eq.~(4) by means of simple algebraic
transformation~(6). The RH problem formalism can be developed equivalently
for either the MNLSE~(2) or Eq.~(4), the RH problem data being invariant
under the gauge transformation. Because the formulation of the RH problem
with the canonical normalization condition has a number of technical
advantages in calculation of soliton solutions, we will develop the RH
formalism for Eq.~(4) and give the transition relations to the MNLSE.

To construct the RH problem associated with Eq.~(4), consider the matrix
Jost-type solutions $J^\p_{\pm}$ of Eq.~(5$a$) which satisfy the asymptotic
conditions
$J^\p_{\pm}\rightarrow\I$ at $\t\rightarrow\pm\infty$. By the standard
analysis of the Volterra-type integral equations for $J^\p_\pm$ which follow
from Eq.~(5$a$) and the above asymptotic properties, we conclude that the
following matrix function ($(J^\p_\pm)_{\cdot l}$ means the $l$-th column of
$J^\p_\pm$)
$$
\Phi^\p_+(k)=\biggl((J^\p_+)_{\cdot 1}(k),\; (J^\p_-)_{\cdot 2}(k)
\biggr),
\eqno(7)$$
being a solution of Eq.~(5$a$), is holomorphic in the two quadrants of the
complex $k$-plane which are defined by the condition $\a$Re$(k)$Im$(k)\le0$.
The scattering matrix $S^\p(k)$ is defined in the usual way:  $$
J^\p_-E=J^\p_+ES^\p, \quad E\equiv \exp\(\L(k)\s\t\).
\eqno(8)$$
Note that det$S^\p=1$ due to det$J^\p_\pm=1$. The Zakharov-Shabat
factorization~[22] of the scattering matrix,
$$
S^\p_+=S^\p S^\p_-,\quad S^\p_+=\(\begin{array}{cc}1&S^\p_{12}\\
0&S^\p_{22}\end{array}\),\quad
S^\p_-=\(\begin{array}{cc}({S^\p}^{-1})_{11}&0\\
({S^\p}^{-1})_{21}&1\end{array}\),
\eqno(9)$$
allows us to represent $\Phi^\p_+(k)$ in two equivalent forms:
$$
\Phi^\p_+=J^\p_+ES^\p_+E^{-1}=J^\p_-ES^\p_-E^{-1}.
\eqno(10)$$
Since the potential $Q^\p$ is Hermitian, we have the following identities:
$$
{J^\p}^\d_\pm(k)={J^\p}^{-1}_\pm(\bar k),\quad
{S^\p}^\d(k)={S^\p}^{-1}(\bar k), \quad
k\in\{k:{\rm Re}(k){\rm Im}(k)=0\}.
$$
Hermiticity of the potential also enables to define the matrix function
conjugated to $\Phi^\p_+(k)$ and holomorphic in the rest two
quadrants of the complex $k$-plane, i.e., which are given by the condition
$\a$Re$(k)$Im$(k)\ge0$:
$$
{\Phi^\p_-}^{-1}(k)={\Phi^\p_+}^\d(\bar k)
=\biggl(({J^\p_+}^{-1})_{1\cdot}(k),\;({J^\p_-}^{-1})_{2\cdot}(k)\biggr)^t,
\eqno(11)$$
where $({J^\p_\pm}^{-1})_{l\cdot}$ denotes the $l$-th row of the matrix
${J^\p_\pm}^{-1}$ and superscript $t$ means transposition. The
linear spectral problem~(5$a$) possesses the parity symmetry~[16]. It can be
summarized to the following important identities:
$$
{\cal P}J^\p_\pm(k)=J^\p_\pm(k), \quad {\cal P}S^\p(k)=S^\p(k), \quad
{\cal P}S^\p_\pm(k)=S^\p_\pm(k),
$$
where ${\cal P}$ is the parity operator defined by ${\cal
P}F(k)=F^{(d)}(-k)-F^{(off)}(-k)\equiv \s F(-k)\s$, $F^{(d)}$
and~$F^{(off)}$ are diagonal and off-diagonal parts of a matrix $F$.
These identities give
$$
{\cal P}\Phi^\p(k)=\Phi^\p(k), \quad \Phi^\p(k)=\(\begin{array}{cc}
\Phi^\p_+(k),& k\in\{k:\a{\rm Re}(k){\rm Im}(k)\le0\},\\
\Phi^\p_-(k),& k\in\{k:\a{\rm Re}(k){\rm Im}(k)\ge0\}.\end{array}\)
\eqno(12)$$
Here $\Phi^\p(k)$ is a matrix function piecewise meromorphic in the
complex $k$-plane and discontinuous through the curve $k\in\{k: {\rm
Re}(k){\rm Im}(k)=0\}$.

Let us return to the gauge equivalence between Eqs.~(2) and~(4). From
Eq.~(12) taken at $k=0$ it follows that the gauge transformation matrix $g$ is
diagonal, while Eqs.~(5$a$) and~(10) lead to the following expression
for~$g$:  $$
g\equiv\Phi^\p(k=0)=\(\begin{array}{cc}\exp\biggl(-i\a\int\limits^\infty_\t{\rm d}\t
|q^\p|^2\biggr)& 0\\ 0& \exp\biggl(-i\a\int\limits^\t_{-\infty}{\rm d}\t |q^\p|^2\biggr)
\end{array}\).
\eqno(13)$$
The matrix function $\Phi(k)$ which results from
$\Phi^\p(k)$ by means of the transformation
$$
\Phi(k)=g^{-1}\Phi^\p(k),
\eqno(14)$$
is a solution of the linear problem~(3$a$), possesses the same conjugation
and meromorphic properties as $\Phi^\p(k)$ does, and is an
eigenfunction of the parity operator~${\cal P}$ too. For the linear
problem~(3$a$), the Jost-type solutions $J_\pm(k)$, the
scattering matrix $S(k)$, and its factorization $S_\pm(k)$ are constructed
similarly to the problem~(5$a$). The matrix function $\Phi(k)$ is given
through the Jost-type solutions and the factorization by the the same
formulas as $\Phi^\p(k)$ does (see Eq.(10)). The following relations are
valid
$$
J_\pm=g^{-1}J^\p_\pm g_\pm, \quad S=g_+^{-1}S^\p g_-, \quad
S_\pm=g_\pm^{-1}S^\p_\pm,
\eqno(15)$$
where $g_\pm=\lim\limits_{\t\rightarrow\pm\infty}g$.

Now we can formulate the RH problem associated with Eq.~(4). Indeed, using
Eqs.~(7) and~(11) as well as the relation $E(\bar k)=E(k)$ for
$k\in\{k: {\rm Re}(k){\rm Im}(k)=0\}$, we write ($\bar
S^\p_{12}(k)\equiv\overline{S^\p_{12}(\bar k)}$)
$$
{\Phi^\p_-}^{-1}(k)\Phi^\p_+(k)=E(k){S^\p_+}^\d(\bar k)S^\p_+(k)E^{-1}(k)
\qquad\qquad\qquad\qquad\qquad$$
$$
\qquad\qquad\qquad\qquad\qquad
=E(k)\(\begin{array}{cc}1& S^\p_{12}(k)\\
\bar S^\p_{12}(k)& 1\end{array}\)E^{-1}(k)\equiv G(k),
\eqno(16a)$$
$$
\Phi^\p(k)\rightarrow\I,\qquad k\rightarrow\infty,
\eqno(16b)$$
where $k\in\{k: {\rm Re}(k){\rm Im}(k)=0\}$.
It is a problem of analytic factorization of the
nondegenerate matrix~$G(k)$ given on the contour defined by the following
disconnected oriented set:
$$
C_\a=\sgn(\a)\biggl(\{(\infty,0),(0,0)\}\cup\{(0,0),(-i\infty,0)\}
\qquad\qquad\qquad$$
$$
\qquad\qquad\qquad
\cup\{(-\infty,0),(0,0)\}\cup\{(0,0),(0,i\infty)\}\biggr)
$$
of the $k$-plane axes. Here sgn$(\a)$ means bypassing the contour in the
reverse direction for~$\a<0$. The functions $\Phi^\p_\pm(k)$ are just a
solution of the RH problem~(16). The uniqueness of this solution is provided
by the canonical normalization condition~(16$b$).

Substituting the asymptotic decomposition of $\Phi^\p(k)$ in the inverse
power series of~$k$,
$$
\Phi^\p(k)=\I+k^{-1}{\Phi^\p}^{[1]}+\ldots,
$$
into the spectral equation~(5$a$) and taking advantage of Eq.~(12), we
reconstruct the potential $Q^\p$:
$$
Q^\p=\frac{1}{\a}[\s,{\Phi^\p}^{[1]}]
=\frac{1}{\a}\lim\limits_{k\rightarrow\infty}k[\s,\Phi^\p(k)].
\eqno(17)$$
Hence, to solve the MNLSE~(2), we should at first solve the RH
problem~(16), then obtain the potential $Q^\p$~(17) and transform it by
Eq.~(6).  It should be noted that the RH problem~(16) remains unchanged
under the gauge transformation~(14), except for the normalization
of the matrix $\Phi(k)$ at infinity, namely, $\Phi(k)\rightarrow g^{-1}$
at~$k\rightarrow\infty$.

In general, the functions det$\Phi^\p_+(k)$ and~det${\Phi^\p_-}^{-1}(k)$ have
zeros in their regions of analiticity, the RH problem being said to be
nonregular (or with zeros). It follows from Eq.~(11) that zeros of the
above determinants are complex conjugate, while the parity symmetry
(Eq.~(12)) tells us that zeros appear by pairs, i.e.,
det$\Phi^\p_+(\pm k_j)=0$ for the $j$-th zero $k_j$.
The solution of the RH problem with zeros can be factorized~[23]
$$
\Phi_\pm^\p(k)=\Phi^\p_{o\pm}(k)\G(k),\quad {\rm det}\Phi^\p_{o\pm}(k)\ne0
\eqno(18)$$
by means of the solution $\Phi^\p_{o\pm}$ of the regular RH problem:
$$
(\Phi^\p_{o-})^{-1}(k)\Phi^\p_{o+}(k)=\G(k)G(k)\G^{-1}(k),
\eqno(19a)$$
$$
\Phi^\p_o(k)\rightarrow\I,\qquad k\rightarrow\infty,
\eqno(19b)$$
which is posed on the same contour~$C_\a$. The matrix function $\G(k)$
represents the contribution of zeros. In the case of the single pair of zeros
$\pm k_1 \equiv k_{\pm1}$ this function is given by (see also Ref.~18)
$$
\G=\I-\sum_{j,l=\pm1}\frac{\vecr{p_j}D^{-1}_{jl}\vecl{p_l}}{k-\bar k_l},
\quad \G^{-1}
=\I+\sum_{j,l=\pm1}\frac{\vecr{p_j}D^{-1}_{jl}\vecl{p_l}}{k-k_j},
\eqno(20)$$
where  $D_{ln}=(k_n-\bar k_l)^{-1}\langle p_l\mid p_n\rangle$,
$D^{-1}_{\cdot\cdot}\equiv (D^{-1})_{\cdot\cdot}$,
$\langle p_l\mid p_n\rangle
=\overline{(p_l)}_1(p_n)_1+\overline{(p_l)}_2(p_n)_2$, vector-rows
$\vecl{p_{\pm1}}$ are related to vector-columns $\vecr{p_{\pm1}}$ by
conjugation, i.e., \mbox{$\vecl{p_{\pm1}}=\vecr{p_{\pm1}}^\d$,} and
the latter are given by $$ \Phi^\p_+(k_{\pm1})\vecr{p_{\pm1}}=0.
\eqno(21)$$
The parity symmetry immediately gives a relation between $\vecr{p_{-1}}$
and~$\vecr{p_{+1}}$. Indeed, the identity
$\Phi^\p_+(k_1)\vecr{p_{+1}}=\s\Phi^\p(k_{-1})\s\vecr{p_{+1}}=0$ leads, due to
uniqueness of~$\vecr{p_{-1}}$, to the relation
$\vecr{p_{-1}}=\s\vecr{p_{+1}}$. From Eq.~(11) it follows that
$\vecl{p_{\pm1}}{\Phi^\p_-}^{-1}(k_{\pm1})=0$. This identity and
Eq.~(21) ensure that the matrix functions $\Phi^\p_{o+}(k)$
and~$(\Phi^\p_{o-})^{-1}(k)$ are holomorphic in the respective quadrants of
the complex $k$-plane.

In the general case of the RH problem with $N$ pairs of zeros of
det$\Phi^\p_+(k)$ the matrix function $\G(k)$ is given by
Eq.~(20) with
$j,l\in\{-N,-N+1,...,-1,1,...N-1,N\}$ and~$k_{-s}\equiv-k_s$.
\bigskip

\noindent
{\bf 3. Soliton solution of MNLSE}
\medskip

The RH problem data are divided into two parts: discrete data
$\{k_j,\\ \vecr{p_{+j}},j=1,...,N\}$ ($2N$ is the whole number of zeros of
det$\Phi^\p_+(k)$) and continuous datum $G(k)$. Soliton solutions correspond
to the RH problem with zeros provided $G(k)=\I$, i.e.,
$\Phi^\p_{o\pm}(k)=\I$. In other words, only the matrix~$\G(k)$ is
responsible for solitons. We will consider the simplest case of
one-soliton solution of the MNLSE~(2). We have from Eqs.~(6) and~(17):
$$
Q=\frac{1}{\a}\G^{-1}(k=0)\lim\limits_{k\rightarrow\infty}k[\s,\G(k)]\G(k=0),
\eqno(22)$$
where the matrices $\G(k)$ and~$\G^{-1}(k)$ are given by Eq.~(18) and it is
taken into account that $g=\G(k=0)$. As regards the coordinate dependence
of the vector-column $\vecr{p_{+1}}$, it is determined by differentiation
of Eq.~(21) with respect to~$\t$ and~$z$ and taking advantage of Eqs.~(5).
It should be noted that Eq.~(21) determines $\vecr{p_{+1}}$ only up to an
arbitrary norm. We obtain in a particular case
$$
\vecr{p_{+1}}_\t=\L(k_1)\s\vecr{p_{+1}},\quad
\vecr{p_{+1}}_z=\O(k_1)\s\vecr{p_{+1}}.
\eqno(23)$$
Integration of the above equations gives
$$
\vecr{p_{+1}}=e^{f\s}\vecr{p^o},\quad \vecr{p^o}=\(\begin{array}{c}
p^o_1\\p^o_2\end{array}\),
$$
where $f=\L(k_1)\t+\O(k_1)z$ and~$\vecr{p^o}$ is an integration constant
determined up to an arbitrary norm. Let us define
$p^o_1/p^o_2=\exp(a+i\pfi)$, where $a$ and~$\pfi$ are real constants.
Ultimately, we have
$$
\vecr{p_{+1}}=\(\begin{array}{c}\exp(a+i\pfi+f)\\e^{-f}\end{array}\),
\quad f=-\frac{2i}{\a}\(k_1^2-\frac{\b}{4}\)\t
-\frac{4i}{\a^2}\(k_1^2-\frac{\b}{4}\)^2z.
\eqno(24)$$
Substituting Eq.~(24) into Eq.~(20), we obtain for $\G(k)$ the following
expression
$$
\G(k)=\I-\frac{2e^a}{k^2-\bar k_1^2}\(\begin{array}{cc}
\bar k_1e^x\(D^{-1}_{++}+D^{-1}_{-+}\)&
ke^{i\psi}\(D^{-1}_{++}+D^{-1}_{-+}\)\\
ke^{-i\psi}\(D^{-1}_{++}-D^{-1}_{-+}\)&
\bar k_1e^{-x}\(D^{-1}_{++}-D^{-1}_{-+}\)\end{array}\).
\eqno(25)$$
Here
$$
x=a+2{\rm Re}f
=a-\frac{8\xi\eta}{\a}\(\t+\frac{4}{\a}\[\xi^2-\eta^2-\frac{\b}{4}\]z\),
$$
$$
\eqno(26)$$
$$
\psi=\pfi+2{\rm Im}f=\pfi
-\frac{4}{\a}\Biggl\{\(\xi^2-\eta^2-\frac{\b}{4}\)\t
+\frac{2}{\a}\Biggl(\[\xi^2-\eta^2-\frac{\b}{4}\]^2-4\xi^2\eta^2\Biggr)z\Biggr\},
$$
and $k_1=\xi-i\eta$ (due to the condition $\a$Re$(k_1)$Im$(k_1)<0$ we will have
$\xi>0$ and~$\eta>0$ for~$\a>0$). From the definition
$D_{ln}=(k_n-\bar k_l)^{-1}\langle p_l\mid p_n\rangle$ we derive the
following properties of the matrix $D$: $D_{++}=-D_{--}$, $D_{-+}=-D_{+-}$.
Hence,
$$
D^{-1}_{++}\pm D^{-1}_{-+}=-\(D_{++}\pm D_{-+}\)({\rm det}D)^{-1}
=\(D_{++}\mp D_{-+}\)^{-1}
$$
$$
=e^{-a}\(\frac{i}{\eta}{\rm ch}x\mp\frac{1}{\xi}{\rm sh}x\)^{-1}.
\eqno(27)$$
Now, we have all to calculate one-soliton solution $q_s$ of the MNLSE~(2).
Substituting Eqs.~(26) and~(27) into Eq.~(25), then, in its turn, Eq.~(25)
into Eq.~(22) we obtain
$$
q_s\equiv Q_{12}=8i\frac{\xi\eta}{\a}\frac{k_1e^{-x}+\bar k_1e^x}
{(k_1e^x+\bar k_1e^{-x})^2}e^{i\psi}.
\eqno(28)$$
This is a general form of the one-soliton solutions to the MNLSE~(2) which
depends on four real parameters $\xi$, $\eta$, $a$, and~$\pfi$. The
solution~(28) is written in terms of the coordinates $x$ and~$\psi$~(26)
comprising linear combinations of the normalized retarded time~$\t$ and
distance along the fiber~$z$. It should be stressed that $\a$
enters the denominator of the soliton solution~(28). In other words, we
account nonperturbatively for the pusle self-steepening effect. Similarly to
the NLSE, the parameters $a$ and~$\psi$ play the role of the initial
position and phase, respectively, while the other two parameters $\xi$
and~$\eta$ do not admit so obvious interpretation. In any case, we see from
Eq.~(28) that the normalized half-width $w$  and velocity $v$ of the soliton are
represented by $$ w=\frac{\a}{2\xi\eta},\quad
v=\frac{\b}{\a}-\frac{4}{\a}(\xi^2-\eta^2).
\eqno(29a)$$
As regards the soliton amplitude $A$, it is natural to admit it in the
following form:
$$ A=\frac{4\xi\eta}{\a(\xi^2+\eta^2)^{\frac{1}{2}}}.
\eqno(29b)$$

The MNLS soliton~(28) has a number of peculiarities which distinct it form
the NLS soliton. First, the MNLS soliton has nonzero phase difference at
its limits. Indeed,
$$
\frac{k_1e^{-x}+\bar k_1e^x}{(k_1e^x+\bar k_1e^{-x})^2}
\longrightarrow\left\{\begin{array}{cc}
{\bar k_1}/{k_1^2},& x\rightarrow\infty,\\
{k_1}/{\bar k_1^2},& x\rightarrow-\infty\end{array}\right.
$$
and the said phase difference reads
$$
\Delta\psi={\rm arg}\(q_s(z\rightarrow\infty)\)
-{\rm arg}\(q_s(z\rightarrow-\infty)\)=6{\rm arg}(k_1).
$$
Further, the integral of the soliton amplitude
$$
\int\limits_{-\infty}^\infty{\rm d}\t|q_s|
=\frac{\pi}{2(\xi^2+\eta^2)^{\frac{1}{2}}}
\sum_{n=0}^{\infty}\(\frac{(\frac{1}{4})_n}{n!}\)^2
\(\frac{2\xi\eta}{\xi^2+\eta^2}\)^{2n},
$$
where $\(\frac{1}{4}\)_n\equiv\(\frac{1}{4}+1\)\(\frac{1}{4}+2\)\cdot...
\cdot\(\frac{1}{4}+n-1\)$, depends on the parameters $\xi$
and~$\eta$, in contrast to the same integral of the NLS soliton which does
not depend on the soliton parameters. Moreover, the important
invariant of Eq.~(2), namely, the number of particles or the optical energy
of the soliton,
$$ E=\int\limits_{-\infty}^\infty{\rm d}\t|q_s|^2
=\frac{4}{\a}{\rm arg}(\bar k_1),\quad 0<|{\rm arg}(\bar k_1)|<\frac{\pi}{2},
$$
has the upper limit $2\pi/|\a|$. The phase difference and
optical energy of the MNLS soliton are related:  $\Delta\psi=-3E\a/2$.

The above properties of the MNLS soliton~(28) resemble those of the dark
NLS soliton, which also has nonzero phase difference and a relation between
its energy and the phase difference~[24]. Nevertheless, the soliton~(28) with $\b>0$
reduces to a bright NLS soliton at $\a\rightarrow0$.
To carry out this limit one should take into account that the Lax pair for
the NLSE should be produced at $\a\rightarrow0$ from the Lax pair~(3) for the
MNLSE. This condition implies that the spectral parameter~$k$ depends
on $\a$ and gives the following prescription:
$2(k^2_{MNLS}-\b/4)/\a\rightarrow-k_{NLS}$ at~$\a\rightarrow0$ or
$$
\frac{2\(\xi^2-\frac{\b}{4}\)}{\a}\rightarrow-\xi_o,
\quad \frac{4\xi\eta}{\a}\rightarrow\eta_o, \quad \a\rightarrow0,
$$
where $k_{MNLS}=\xi-i\eta$ and~$k_{NLS}=\xi_o+i\eta_o$
In other words, we have the decomposition
$$
\xi=\frac{\sqrt{\b}}{2}-\frac{\a}{2\sqrt{\b}}\,\xi_o+O(\a^2), \quad
\eta=\frac{\a}{2\sqrt{\b}}\,\eta_o+O(\a^2), \quad \a\rightarrow0,
\eqno(30)$$
which transforms the MNLS soliton~(28) to the NLS soliton
\break
$q_s=(2i\eta_o/\sqrt{\b})e^{i\psi_o}{\rm sech}x_o$, where
$\psi_o=\pfi+2\xi_o\t-2(\xi_o^2-\eta_o^2)z$, $x_o=a-2\eta_o(\t-2\xi_o z)$.
\bigskip

\noindent
{\bf 4. Perturbation-induced evolution of RH problem data}
\medskip

Following our strategy, to find correction to the soliton solution of the
MNLSE caused by a perturbation, we should at first derive the
perturbation-induced evolution equations for the RH problem data.
Decomposition of these equations in the asymptotic power series with
respect to the perturbation will produce the consequent corrections to the
soliton solution.

In Sec.~II we have seen that the gauge equivalent Eqs.~(2)
and~(4) have resembling IST schemes with simple mutual relations~(15).
As the RH problem data are invariant under the gauge transformation~(6),
we can choose between the two IST formulations the most convenient one for
calculation of corrections to soliton solutions. Consider eq.~(2) with a
small perturbation at the r.h.s. of the equation, i.e., Eq.~(1). The
perturbation causes a variation $\delta U$ of the potential, what, in its
turn, leads to a variation $\delta J_\pm$ of the Jost-type solutions. From
the spectral problem~(3$a$) we  obtain the following equation for $\delta
J_\pm$ $$ \(\delta J_\pm\)_\t=-\L(k)\delta J_\pm\s+\delta UJ_\pm+U\delta
J_\pm, \quad \lim\limits_{\t\rightarrow\pm\infty}\(J_\pm^{-1}\delta
J_\pm\)=0.  $$ Therefore, $$ \delta J_\pm=J_\pm
E\(\,\int\limits^\t_{\pm\infty}{\rm d}\t E^{-1}J_\pm^{-1}\delta UJ_\pm
E\,\)E^{-1}, \eqno(31)$$ where $\delta U=\(\delta U/\delta z\)\delta z$. It
should be stressed that the same representation of the variation $\delta
J^\p_\pm$ of the Jost-type solutions to the spectral problem~(5$a$) follows
from Eq.~(31) by the simple substitutions $U\rightarrow U^\p$
and~$J_\pm\rightarrow J^\p_\pm$ and such a procedure can be carried out at
any step below.

Now we introduce a useful matrix function
$$
\g(\pm\infty,\t)=\int\limits^\t_{\pm\infty}{\rm d}\t
E^{-1}\Phi_+^{-1}\frac{\delta U}{\delta z}\Phi_+E.
\eqno(32)$$
Then the variation derivative $\delta J_\pm/\delta z$ takes the
form
$$
\frac{\delta J_\pm}{\delta z}=J_\pm ES_\pm\g(\pm\infty,\t)S_\pm^{-1}E^{-1}.
\eqno(33)$$
From the evident relation $S=\lim\limits_{\t\rightarrow\infty}E^{-1}J_-E$
we obtain the variation of the scattering matrix:
$$
\frac{\delta S}{\delta z}=S_+\g(-\infty,\infty)S_-^{-1},
\eqno(34)$$
as well as of its factorization~(9):
$$
\frac{\delta S_+}{\delta z}=S_+\g(-\infty,\infty)M_{22}, \quad
M_{22}=\(\begin{array}{cc}0&0\\0&1\end{array}\).
\eqno(35)$$
What regards the variation $\delta\Phi_+(k)$
for~$k\in\{$Re$(k)$Im$(k)=0\}$, we obtain from Eqs.~(10), (33), and~(35):
$$
\frac{\delta\Phi_+}{\delta z}=\frac{\delta}{\delta z}
\(J_+ES_+E^{-1}\)=\Phi_+E\biggl(\g(\infty,\t)+\g(-\infty,\infty)M_{22}
\biggr)E^{-1}
$$
$$
=\Phi_+E\biggl(-\g(\t,\infty)M_{11}+\g(-\infty,\t)M_{22}\biggr)E^{-1},
$$
where $M_{11}=\;$diag$(1,0)$. Hence, in the case of perturbation
the evolution equation for~$\Phi_+$ with respect to~$z$~(3$a$) gains an
additional term responsible for the perturbation:
$$
(\Phi_+)_z=V\Phi_+-\O(k)\Phi_+\s+\Phi_+E\Pi E^{-1},
\eqno(36)$$
where the fundamental matrix function $\Pi(k)$ is defined as
$$
\Pi(k)=-\g(\t,\infty)M_{11}+\g(-\infty,\t)M_{22}=\(\begin{array}{cc}
-\g_{11}(\t,\infty)& \g_{12}(-\infty,\t)\\
-\g_{21}(\t,\infty)& \g_{22}(-\infty,\t)\end{array}\).
\eqno(37)$$

Eq.~(36) was derived for~$k$ belonging to the curve
$\{$Re$(k)$Im$(k)\}=0$, involved in the formulation of the RH problem,
but it can be analytically continued into the domain
$\{k:\a{\rm Re}(k){\rm Im}(k)\le0\}$, where $\Phi_+(k)$ is holomorphic.
Indeed, from the linear problem~(3$a$) we get
$$
\(\Phi_+^{-1}\delta\Phi_+\)_\t=\Phi_+^{-1}\delta U\Phi_+
+\L(k)[\s,\Phi_+^{-1}\delta\Phi_+].
$$
Hence, the integrand of Eq.~(32) is represented as
$\(E^{-1}\Phi_+^{-1}(\delta\Phi_+/\delta z)E\)_\t$ and the matrix
function $\g(\pm\infty,\t)$ reads
$$
\g(\pm\infty,\t)=E^{-1}\Phi_+^{-1}\frac{\delta\Phi_+}{\delta z}E
-S_\pm^{-1}\frac{\delta S_\pm}{\delta z},
\eqno(38)$$
where we have used Eq.~(10) to compute the limits $\t\rightarrow\pm\infty$.
Now, in virtue of the explicit structure of the matrices $S_\pm$~(9) and
analiticity of their diagonal elements in the domain
$\{k:\a{\rm Re}(k){\rm Im}(k)\le0\}$, we conclude that the entries of the
matrix $\Pi(k)$~(37) are functions meromorphic in this domain and having simple
poles at the zeros of the RH problem, i.e., at the zeros of det$\Phi_+(k)$.
Analiticity of $\Phi_+(k)$ in the above domain gives immmediately an
important identity
$$
\Phi_+(\pm k_j)E(\pm k_j){\rm Res}\{\Pi(k),\pm k_j\}=0,
\eqno(39)$$
where Res$\{\cdot,\pm k_j\}$ stands for the residue at $k=\pm k_j$,
det$\Phi_+(\pm k_j)=0$.

In order to derive the perturbation-induced evolution of the RH problem
data, let us consider for simplicity the case of a single pair of zeros
$\pm k_1$. The full set of the independent data involves in this case
$k_1$, $\vecr{p_{+1}}$, and~$G(k)$. By the definition of $k_1$, we have
$$
0=\frac{{\rm d}}{{\rm d}z}{\rm det}\Phi_+(k_1)=\frac{\partial}{\partial z}
{\rm det}\Phi_+(k_1)+\frac{{\rm d}k_1}{{\rm d}z}
\(\frac{\partial}{\partial k} {\rm det}\Phi_+\)_{k=k_1}
$$
$$
=\biggl({\rm tr}\Pi(k){\rm det}\Phi_+(k)\biggr)_{{k=k_1}}+\frac{{\rm
d}k_1}{{\rm d}z} \(\frac{\partial}{\partial k} {\rm
det}\Phi_+\)_{{k=k_1}}.
$$
Here we took advantage of Eq.~(36) and of the
identity tr$V=0$.  Decomposition of $\Pi(k)$ into the regular and
singular parts at $k=k_1$, $$ \Pi(k)=\Pi_r(k)+\frac{{\rm
Res}\{\Pi(k),k_1\}}{k-k_1}\,, \eqno(40)$$ and use of the formula
det$\Phi_+={\rm det}\Phi_{o+}{\rm det}\G ={\rm
det}\Phi_{o+}(k^2-k_1^2)(k^2-\bar k_1^2)^{-1}$ which follows form
Eqs.~(14), (18), and~(20) with $\Phi_{o}\equiv g^{-1}\Phi_o^\p$ help us
to compute the limit $k\rightarrow k_1$:
$$
\frac{{\rm d}k_1}{{\rm d}z}=\!-\!\(\frac{{\rm tr}\[\Pi_r(k)
\!+\!(k\!-\!k_1)^{-1}{\rm Res}\{\Pi(k),k_1\}\](k^2\!-\!k_1^2)(k^2\!-\!\bar k_1^2)^{-1}
{\rm det}\Phi_{o+}(k)}{\frac{\partial}{\partial k}
\[(k^2\!-\!k_1^2)(k^2\!-\!\bar k_1^2)^{-1}{\rm
det}\Phi_{o+}(k)\]}\)_{{k=k_1}}
$$
$$ =-{\rm Res}\{{\rm tr}\Pi(k),k_1\}.
\eqno(41)$$
The perturbation-induced evolution equation for the vector $\vecr{p_{+1}}$
can be obtained by differentiation of Eq.~(21):
$$
\(\frac{{\rm d}}{{\rm d}z}\Phi_+(k)\)_{{k=k_1}}\!\!\vecr{p_{+1}}
+\Phi_+(k_1) \frac{{\rm d}}{{\rm d}z}\vecr{p_{+1}}=0.
\eqno(42)$$
Here the full derivative of $\Phi_+(k)$
with respect to~$z$ is given, due to Eq.~(36), as follows:
$$
\(\frac{{\rm d}}{{\rm d}z}\Phi_+(k)\)_{{k=k_1}}\!\vecr{p_{+1}}
=\(\frac{\partial}{\partial z}\Phi_+(k)\)_{{k=k_1}}\!\vecr{p_{+1}}
+ \frac{{\rm d}k_1}{{\rm d}z}\(\frac{\partial}{\partial k}
\Phi_+(k)\)_{{k=k_1}}\!\!\vecr{p_{+1}} \qquad\quad
$$
$$ =\[-\O(k)\Phi_+(k)\s
\!+\!\Phi_+(k)E(k)\(\Pi_r(k)\!+\!\frac{{\rm Res}\{\Pi(k),k_1\}}{k-k_1}\)E^{-1}(k)
\]_{{k=k_1}}\!\vecr{p_{+1}}
$$
$$
+\frac{{\rm d}k_1}{{\rm d}z}\(\frac{\partial}{\partial
k}\Phi_+(k)\)_{{k=k_1}}\!\!\vecr{p_{+1}},
\eqno(43)$$
where we have used the identity~(21) to exclude an evidently vanishing term:
$V(k_1)\Phi_+(k_1)\vecr{p_{+1}}=0$.  Turning again to Eqs.~(36) and~(40),
we obtain the following relation:
$$
\[(k-k_1)\Phi_+^{-1}(k)(\Phi_+)_z(k)\]_{{k=k_1}}\!\!\vecr{p_{+1}}
=E(k_1){\rm Res}\{\Pi(k),k_1\}E^{-1}(k_1)\vecr{p_{+1}}.
$$
On the other hand, the l.h.s. of this relation can be transformed in that
way:
$$
\[\(k-k_1\)\Phi_+^{-1}(k)(\Phi_+)_z(k)\]_{{k=k_1}}\!\!\vecr{p_{+1}}
=-\frac{{\rm d}k_1}{{\rm d}z}\vecr{p_{+1}}
\quad\qquad\qquad\qquad\qquad\qquad$$
$$
\quad\qquad\qquad\qquad\qquad
-\(\[\(k-k_1\)\Phi_+^{-1}(k)\]_z\)_{{k=k_1}}\Phi_+(k_1)
\vecr{p_{+1}}
=-\frac{{\rm d}k_1}{{\rm d}z}\vecr{p_{+1}},
$$
where we have once again used Eq.~(21). Hence, we have derived an
important identity $$ E(k_1){\rm Res}\{\Pi(k),k_1\}E^{-1}(k_1)\vecr{p_{+1}}=
-\frac{{\rm d}k_1}{{\rm d}z}\vecr{p_{+1}}.
\eqno(44)$$
Using now Eq.~(39) we can write
$$
\[\Phi_+(k)E(k)\frac{{\rm Res}\{\Pi(k),k_1\}}{k-k_1}
E^{-1}(k)\]_{{k=k_1}}\!\!\vecr{p_{+1}}
\quad\qquad\qquad\qquad\qquad\qquad\qquad$$
$$
\qquad\qquad\qquad\qquad
=\(\frac{\partial}{\partial k}\Phi_+(k)\)_{{k=k_1}}E(k_1)
{\rm Res}\{\Pi(k),k_1\}E^{-1}(k_1)\vecr{p_{+1}}.
\eqno(45)$$
Collecting Eqs.~(42)-(45), we obtain the perturbation-induced evolution of
the vector $\vecr{p_{+1}}$,
$$
\frac{{\rm d}}{{\rm d}z}\vecr{p_{+1}}=\O(k_1)\s\vecr{p_{+1}}
-E(k_1)\Pi_r(k_1)E^{-1}(k_1)\vecr{p_{+1}},
$$
or, for $\t$-independent vector
$\vecr{p^o}\equiv F^{-1}(k_1)E^{-1}(k_1)\vecr{p_{+1}}$, where
$F(k)\equiv\exp\(\int\limits^z{\rm d}z\O(k)\s\)$,
$$
\frac{{\rm d}}{{\rm d}z}\vecr{p^o}=-F^{-1}(k_1)\Pi_r(k_1)F(k_1)\vecr{p^o}.
\eqno(46)$$
Finally, we should derive an evolution equation for the
continuous datum~$G(k)$.  Eqs.~(3$b$), (11), (16) (written for the
quantities without prime), and~(36) give $$
G_z(k)=\O(k)[\s,G(k)]+G(k)E(k)\Pi(k)E^{-1}(k)
+E(k)\Pi^\d(\bar k)E^{-1}(k)G(k).
$$
This equation is simplified by introducing the $\t$-independent
matrix~$G^o\equiv F^{-1}E^{-1}GEF$:
$$
G^o_z(k)=G^o(k)F^{-1}(k)\Pi(k)F(k)+F^{-1}(k)\Pi^\d(\bar k)F(k)G^o(k).
\eqno(47)$$
It is evident that $\t$-independence of the l.h.s. of Eqs.~(41), (46),
and~(47) enables us to simplify further these equations by taking one of
the limits $\t\rightarrow\pm\infty$, at which the fundamental matrix
$\Pi(k)$~(37) has only one nonzero column.

As was pointed above, we can use the gauge equivalent IST schemes
for the derivation of perturbation-induced evolution equations for the RH
problem data, due to the gauge invariance of the latter.  Indeed, from
Eqs.~(38), (14), and~(15), using the parity symmetry, one can easily find the
folowing relation between the gauge equivalent fundamental matrices:
$$
\Pi^\p(k)=\Pi(k)-E^{-1}(k)\Phi^{-1}_+(k)
\Pi^{(d)}(k\rightarrow\infty)\Phi_+(k)E(k)
\qquad\qquad\qquad$$
$$
\qquad= \Pi(k)+E^{-1}(k)\Phi^{-1}_+(k)\(\begin{array}{cc}
\g_{11}(\t,\infty)&0\\0&-\g_{22}(-\infty,\t)\end{array}\)_{k\rightarrow\infty}
\!\!\Phi_+(k)E(k).
$$
where $\Pi^\p(k)$ is associated with the canonical equation~(4). It is easy
to see that the second term in the r.h.s. of this expression for~$\Pi^\p(k)$
gives no contribution to the evolution equations~(41), (46), and~(47). For
instance, Res$\{{\rm tr}\Pi^\p(k),k_1\}=\;$Res$\{{\rm tr}\Pi(k),k_1\}$.
This once again ensures that one can use either of the gauge equivalent
fundamental matrices $\Pi$ or~$\Pi^\p$ expressed through $\Phi_+$, $\delta
U/\delta z$ or~$\Phi^\p_+$, $\delta U^\p/\delta z$, respectively, for
calculation of corrections.  It will be more convenient to use the
former fundamental matrix~$\Pi$, because precisely
$\delta U_{12}/\delta z$ is proportional to the r.h.s. of Eq.~(1), treated
as the perturbation to the MNLSE.
\bigskip

\noindent
{\bf 5. Adiabatic approximation}
\medskip

It should be stressed that the evolution equations (41), (46), and~(47) of
the preceding section are exact, but just this circumstance prevents their
direct use for concrete calculations. They include unknown function
$\Phi_+(k)$, a solution of the perturbed equation~(36). The way to
determine $\Phi_+(k)$ explicitly defines the sort of the approximation used.
Here we develop the simplest variant to account for perturbation, the so
called adiabatic approximation. In the adiabatic approximation, soliton
shape is considered to be unchanged instantaneously under the action of a
small perturbation, while the soliton parameters, being constant in the
integrable case, acquire slow $z$-dependence. In other words, we impose the
condition $G(k)=\I$ and, consequently, $\Phi^\p_{o+}(k)=\I$~(19$a$). Then,
Eqs.~(14), (13) and~(18) give
\mbox{$\Phi_+(k)=g^{-1}\Phi^\p_+(k)=\G^{-1}(k=0)\G(k)$}.  Let us define the
perturbation matrix~$R$ as
$$ \frac{\delta U}{\delta z}=2ik\frac{\delta
Q}{\delta z}=2kR.
\eqno(48)$$
Then the r.h.s. of Eq.~(1) is nothing but~$R_{12}$. As was mentioned above,
taking the limit $\t\rightarrow\pm\infty$ considerably simplifies the
evolution equations~(41) and~(46) for soliton parameters.  From Eq.~(32) we
obtain $$
\g(k)\equiv\g(\!-\infty,\!\infty)=2k\!\int\limits^{\infty}_{-\infty}\!{\rm
d}\t E^{-1}(k)\G{-1}(k)\G(k\!=\!0)R\G{-1}(k\!=\!0)\G(k) E(k).
\eqno(49)$$
At the limit $\t\rightarrow\infty$ the fundamental matrix~$\Pi(k)$ defined
by Eq.~(37) takes the form
$$
\Pi_{\t\rightarrow\infty}(k)=\(\begin{array}{cc}0&\g_{12}(k)\\
0&\g_{22}(k)\end{array}\)_{\t\rightarrow\infty}.
\eqno(50)$$
With this representation for~$\Pi(k)$, the evolution equation~(41) is
written as ${\rm d}k_1/{\rm d}\t=-{\rm Res}\{\g_{22}(k),k_1\}$.
Taking into account Eqs.~(25) and~(27), from Eqs.~(41) and~(49) we obtain
$$
\frac{{\rm d}k_1}{{\rm d}z}=i\a k_1^2\int\limits^\infty_{-\infty}{\rm d}xe^x
\frac{r_o(x)+\bar r_o(-x)}{\(k_1e^{-x}+\bar k_1e^x\)^2},
\eqno(51)$$
where $r_o(x,z)\equiv e^{-i\psi(x,z)}r(x,z)$ and the variables $x$
and~$\psi$ are given by Eqs.~(26) with account of possible $z$-dependence of
the soliton parameters $\xi$ and~$\eta$:  $$ x=a-\frac{8\xi\eta}{\a}\t
-\frac{32}{\a^2}\int\limits^z{\rm d}z\,\xi\eta\(\xi^2-\eta^2-\frac{\b}{4}\),
$$
$$
\psi=\pfi-\frac{4}{\a}\left\{\(\xi^2-\eta^2-\frac{\b}{4}\)\t
+\frac{2}{\a}\int\limits^z{\rm d}z
\(\[\xi^2-\eta^2-\frac{\b}{4}\]^2-4\xi^2\eta^2\)\right\}.
$$
For the soliton parameters $\xi$ and~$\eta$, $k_1=\xi-i\eta$, we get from
Eq.~(51):
$$
\frac{{\rm d}\xi}{{\rm d}z}=\frac{i\a}{2}\int\limits^\infty_{-\infty}{\rm d}x
\frac{k_1^2e^x-\bar k_1^2e^{-x}}{\(k_1e^{-x}+\bar k_1e^x\)^2}
\(r_o(x)+\bar r_o(-x)\),
\eqno(52)$$
$$
\frac{{\rm d}\eta}{{\rm d}z}=-\frac{\a}{2}\int\limits^\infty_{-\infty}
{\rm d}x\frac{k_1^2e^x+\bar k_1^2e^{-x}}{\(k_1e^{-x}+\bar k_1e^x\)^2}
\(r_o(x)+\bar r_o(-x)\).
\eqno(53)$$
It is noteworthy that only the symmetric part of the perturbation, i.e.,
such that $\bar r_o(-x,z)=r_o(x,z)$, changes the soliton parameters $\xi$
and~$\eta$.

In order to obtain the adiabatic evolution for the rest two soliton
parameters $a$ and~$\pfi$, we turn to Eq.~(46). Rewriting it by entries
with the matrix $\Pi$ given by Eq.~(50) and taking into account Eq.~(40),
we obtain
$$
\frac{{\rm d}p^o_1}{{\rm d}z}=-\exp\biggl(-2\int\limits^z{\rm d}z
\O(k_1)\biggr){\g_r}_{12}(k_1)\,p^o_2,\quad
\frac{{\rm d}p^o_2}{{\rm d}z}=-{\g_r}_{22}(k_1)\,p^o_2
$$
and for $p^o_1/p^o_2\equiv e^{a+i\pfi}$:
$$
\frac{{\rm d}}{{\rm d}z}e^{a+i\pfi}=e^{a+i\pfi}{\g_r}_{22}(k_1)
-\exp\biggl(-2\int\limits^z{\rm d}z\O(k_1)\biggr){\g_r}_{12}(k_1).
\eqno(54)$$
Now we should extract the regular part of~$\g(k)$ according to the
decomposition $\g_{\cdot2}(k)={\g_r}_{\cdot2}(k)+{\rm Res}
\{\g_{\cdot2}(k),k_1\}(k-k_1)^{-1}$~(40). After long but not difficult calculations
with account of the explicit expression for $\G(k)$~(25) we get $\(C\equiv
a-(32/\a^2)\int\limits^z{\rm d}z\,\xi\eta\(\xi^2-\eta^2-{\b}/{4}\)\)$
$$
{\g_r}_{12}(k_1)=\frac{\a k_1}{2\xi\eta}
\exp\biggl(2\int\limits^z{\rm d}z\O(k_1)\biggr)e^{a+i\pfi}
\int\limits^\infty_{-\infty}\frac{{\rm d}x}{\(k_1e^{-x}+\bar k_1e^x\)^2}
\qquad\qquad\qquad\qquad$$
$$
\times
\biggl\{\biggl[\(2k_1^2C-3i\xi\eta\)\(r_o(x)+\bar r_o(-x)\)
+\(\xi^2-\eta^2-2k_1^2x\)
\(r_o(x)-\bar r_o(-x)\)\biggr]e^x
$$
$$
+2|k_1|^2r_o(x)e^{-x}\biggr\},
$$
$$
{\g_r}_{22}(k_1)=\frac{\a k_1}{2\xi\eta}
\int\limits^\infty_{-\infty}{\rm d}x
\frac{r_o(x)+\bar r_o(-x)}{\(k_1e^{-x}+\bar k_1e^x\)^2}
\(|k_1|^2e^{-x}-3i\xi\eta e^x\).
\qquad\qquad\qquad\qquad$$
Now from Eq.~(54) we obtain
$$
\frac{{\rm d}a}{{\rm d}z}=-\frac{\a}{2\xi\eta}
\int\limits^\infty_{-\infty}\frac{{\rm d}x}{\(k_1e^{-x}+\bar k_1e^x\)^2}
\biggr\{C\(k_1^3e^x+\bar k_1^3e^{-x}\)\(r_o(x)+\bar r_o(-x)\)
\qquad\qquad\qquad$$
$$
-\biggl[i\xi\eta\(k_1e^x+\bar k_1e^{-x}\)+x\(k_1^3e^x+\bar k_1^3e^{-x}\)
\biggr]\(r_o(x)-\bar r_o(-x)\)\biggr\},
\eqno(55)$$
$$
\frac{{\rm d}\pfi}{{\rm d}z}=\frac{i\a}{2\xi\eta}
\int\limits^\infty_{-\infty}\frac{{\rm d}x}{\(k_1e^{-x}+\bar k_1e^x\)^2}
\biggr\{C\(k_1^3e^x-\bar k_1^3e^{-x}\)\(r_o(x)+\bar r_o(-x)\)
\qquad\qquad\qquad$$
$$
\biggl[k_1e^x\(\bar k_1^2-i\xi\eta\)+\bar k_1e^{-x}\(k_1^2+i\xi\eta\)
-x\(k_1^3e^x-\bar k_1^3e^{-x}\)\biggr]\(r_o(x)-\bar r_o(-x)\)\biggr\}.
\eqno(56)$$

\noindent
For the case of the symmetric perturbation, i.e., when $\bar
r_o(-x,z)=r_o(x,z)$ holds, the evolution of~$a$ and~$\pfi$ is
determined by the evolution of~$k_1$:
$$
\frac{{\rm d}a}{{\rm d}z}=-\frac{C}{\xi\eta}
{\rm Im }\left\{k_1\frac{{\rm d}k_1}{{\rm d}z}\right\},\quad
\frac{{\rm d}\pfi}{{\rm d}z}=\frac{C}{\xi\eta}
{\rm Re}\left\{k_1\frac{{\rm d}k_1}{{\rm d}z}\right\}.
\eqno(57)$$
Hence, in the symmetric case the adiabatic evolution of the soliton
parameters is determined by a single integral involved in the r.h.s. of
Eq.~(51). It is remarkable that the symmetry condition $\bar
r_o(-x,z)=r_o(x,z)$ is rather natural for the MNLSE, because the
soliton~(28) itself satisfies this condition:
$\overline{q_s(-x,z)e^{-i\psi}}=q_s(x,z)e^{-i\psi}$.
\bigskip

\noindent
{\it Example}

As an example of using the above formalism, we analyse here in the
adiabatic approximation the effect of two different perturbations to the
soliton propagation. We will consider a perturbation comprising two terms,
namely, the linear perturbation which describes the fiber loss or
excess linear gain and the Raman driving which is reponsible for the
soliton self-frequency shift effect~[19-21]. Such a choice of perturbation
types does not in any way exhaust the set of all perturbations relevant for
the femtosecond soliton propagation in optical
fibers and serves mainly as an illustration of the
perturbation theory developed above . The results will be presented in a
form suitable for numerical calculations.

Let us write the perturbation in
the following form:
$$ r=i\eps_1q+\eps_2|q|^2_\t q,
\eqno(58)$$
where
$\eps_1$ and~$\eps_2$ are small parameters characterizing the linear gain
($\eps_1>0$) or the fiber losses ($\eps_1<0$) and the Raman driving,
respectively. Substituting the soliton solution~(28) into Eq.~(58), we
obtain the characteristic perturbation $r_o$ in the form
$$
r_o=-4\eps_1\frac{\xi\eta}{\a}\frac{k_1e^{-x}+\bar k_1e^x}
{(k_1e^x+\bar k_1e^{-x})^2}
\qquad\qquad\qquad\qquad\qquad\qquad$$
$$
\qquad\qquad\qquad+i\eps_2\(\frac{8\xi\eta}{\a}\)^4
\frac{e^{2x}-e^{-2x}}{(k_1e^x+\bar k_1e^{-x})^4\(k_1e^{-x}+\bar k_1e^x\)}.
\eqno(59)$$

\noindent
Note that this perturbation is symmetric in the sense of the preceding
section, i.e., $\bar r_o(-x,z)=r_o(x,z)$. Hence, evolution of the
soliton parameters $a$ and~$\pfi$ is determined by evolution of~$k_1$ in
accordance with Eq.~(57). With the perturbation~(59) Eq.~(51) takes
surprizingly simple form:
$$
\frac{{\rm d}k_1}{{\rm d}z}=-\eps_1\frac{k_1}{2}\(1-2\th{\rm
cot}(2\th)-2i\th\)
\qquad\qquad\qquad\qquad\qquad\qquad\qquad$$
$$
\qquad\qquad\qquad-16\eps_2\frac{k_1|k_1|^4}{\a^3}\(1-2\th{\rm cot}(2\th)
-\frac{1}{3}{\rm sin}^2(2\th)\),
\eqno(60)$$
with
$$
\frac{{\rm d}\th}{{\rm d}z}=2\eps_1\th.
\eqno(61)$$
Here $\th\equiv\;$arctan$(\eta/\xi)$, $0<\th<\pi/2$,
cot$(2\th)=(\xi^2-\eta^2)/(2\xi\eta)$,
and~sin$(2\th)=2\xi\eta/(\xi^2+\eta^2)$.

It should be noted that the parameter $\a$ enters Eq.~(60) in a denominator
representing thereby a nonperturbative account of selfsteepening of the
soliton. In the limit~(30) Eq.~(60) reduces to the well-known equations~[5]
for the NLS soliton parameters. From Eq.~(60) one can derive evolution
equations for the soliton parameters~(29):
$$
\frac{{\rm d}w^{-1}}{{\rm d}z}=-\[\frac{\eps_1}{2}(1\!-\!2\th{\rm cot}(2\th))
+32\eps_2\frac{k_1|k_1|^4}{\a^3}\(1\!-\!2\th{\rm cot}(2\th)
\!-\!\frac{1}{3}{\rm sin}^2(2\th)\)\]w^{-1}
$$
$$
+\frac{\eps_1}{2}\th\(v-\frac{\b}{\a}\),
$$

$$
\frac{{\rm d}v}{{\rm d}z}=-\[\eps_1(1\!-\!2\th{\rm cot}(2\th))
+32\eps_2\frac{k_1|k_1|^4}{\a^3}\(1\!-\!2\th{\rm cot}(2\th)
\!-\!\frac{1}{3}{\rm sin}^2(2\th)\)\]\(v-\frac{\b}{\a}\)
$$
$$
-8\eps_1\th w^{-1},
$$

$$
\frac{{\rm d}A}{{\rm d}z}=\[\frac{\eps_1}{2}(6\th{\rm cot}(2\th)\!-\!1)
-16\eps_2\frac{|k_1|^4}{\a^3}\(1-2\th{\rm cot}(2\th)
\!-\!\frac{1}{3}{\rm sin}^2(2\th)\)\]A.
$$
It follows from Eqs.~(61) that the soliton energy
$E=\int\limits_{-\infty}^{\infty}{\rm d}\t|q|^2=4\th/|\a|$
does not depend on the Raman driving:
$$
\frac{{\rm d}E}{{\rm d}z}=2\eps_1E,
$$
just as in the case of the NLSE~[5].
\bigskip

\noindent
{\bf 6. Conclusion}
\medskip

We have developed a perturbation theory for the MNLS soliton. It has been
shown recently~[25] that the MNLSE utilizing the notion of slowly varying
envelope is still valid up to 3 - 5 periods of field oscillations within the
envelope. Our formalism is based on the RH problem associated with the
Wadati-Konno-Ichikawa linear spectral problem.  We have demonstrated
previously the efficiency of the similar approach to the perturbed Manakov
system~[15], as well as to nonlinear evolution equations integrable by the
$N\times N$ Zakharov-Shabat spectral problem~[14]. It is not mere a chance
that the formalisms for both linear and quadratic spectral
bundles resemble each
other. Indeed, the main idea behind the RH-based approach consists in the
analysis of analytical behavior of the Jost-type solutions to the spectral
problem. Besides, we have stressed the gauge invariance of the RH problem
data. Just this property is responsible for the definite unification of the
objects used in both cases. In the recent paper~[26] the RH problem has been
used for studying asymptotics of the MNLSE solution associated with the
continuous spectral data.

As regards the examples of perturbations considered above, we confine
ourselves to the adiabatic approximation. This restriction is by no means
concerned with the present approach. Moreover, we have derived the
perturbation-induced evolution equation~(47) for the continuous RH problem
datum. It is precisely this equation which is necessary for the description
of the soliton shape distortion and emission of linear waves. This problem
will be solved in a forthcoming paper.
\bigskip

\noindent
{\bf Acknowledgements}
\medskip

The authors wish to thank Yu.S. Kivshar for sending a copy of Ref. [24] and
 S.Yu. Sakovich for evaluation of some integrals
with "Mathematica". The research of V.S. was partially supported by the
Foundation for Fundamental Research of the Republic of Belarus,
Grant No. M$\Pi$96-06.\\


\end{document}